\documentstyle[12pt] {article}
\newcommand{\beq}{\begin{equation}}
\newcommand{\eeq}{\end{equation}}
\newcommand{\beqa}{\begin{eqnarray}}
\newcommand{\eeqa}{\end{eqnarray}}
\newcommand{\ba}{\begin{array}}
\newcommand{\ea}{\end{array}}

\begin{document}

\begin{flushright}
To be published in \\
Modern Physics Letters B (1998)
\end{flushright}

\begin{center}
{\large \bf THE ROLE OF DIMENSIONALITY \\
IN THE STABILITY OF \\
A CONFINED CONDENSED BOSE GAS}
\end{center}

\vskip 1. truecm

\begin{center}
{\bf Luca Salasnich}
\footnote{E--Mail: salasnich@padova.infn.it}
\vskip 0.5 truecm
Dipartimento di Matematica Pura ed Applicata,\\
Universit\`a di Padova, Via Belzoni 7, I 35131 Padova, Italy\\
Istituto Nazionale di Fisica Nucleare, Sezione di Padova, \\
Via Marzolo 8, I 35131 Padova, Italy \\
Dipartimento di Fisica, Universit\'a di Milano, and \\
Istututo Nazionale per la Fisica della Materia, Unit\'a di Milano, \\
Via Celoria 16, 20133 Milano, Italy

\end{center}

\newpage

\begin{center}
{\bf Abstract}
\end{center}

\vskip 0.4 truecm
\par
We study analytically the ground--state 
stability of a Bose--Einstein condensate (BEC) confined 
in an harmonic trap with repulsive or attractive zero--range interaction 
by minimizing the energy functional of the system. 
In the case of repulsive interaction the BEC mean radius 
grows by increasing the number of bosons, 
instead in the case of attractive interaction 
the BEC mean radius decreases by increasing the number of bosons:  
to zero if the system is one--dimensional and 
to a minimum radius, with a maximum number of bosons, 
if the system is three--dimensional. 

 
\newpage

\par
In the last three years there has been a renewed interest in the 
Bose-Einstein condensation due to the spectacular experiments 
with alkali vapors $^{87}Rb$, $^{23}Na$ and $^{7}Li$ 
confined in magnetic traps and cooled down to a temperature of the order of 
$100$ nK$^{1),2),3)}$. Numerical studies of the Bose--Einstein condensate (BEC) 
have been performed for the ground state$^{4),5),6)}$ and the 
collective excitations$^{7),8),9)}$. 
\par
In the present paper we analyze analytically 
the ground--state stability of the BEC by minimizing the energy functional 
with respect to the standard deviation of a Gaussian trial wave--function. 
Our analytical results are in good agreement with the numerical calculations 
of Edward and Burnett$^{5)}$ and also of Dalfovo and Stringari$^{6)}$. 
We are able to estimate the maximum number of bosons for which the 
BEC is stable. Moreover we find strong differences 
between the one--dimensional and the three--dimensional cases.
\par
For the alkali vapors 
the range of the atom--atom interaction is believed to be short in 
comparison to the typical length scale of variations of atomic wave functions. 
The atom--atom interaction can be replaced by an effective 
zero--range interaction potential$^{5)}$
\beq
U({\bf r}-{\bf r}')=B\delta ({\bf r}-{\bf r}') \; . 
\eeq
Such an effective potential leads automatically to s--wave scattering 
only, where $B={2\pi \hbar^2 a/m}$ 
is the scattering amplitude and $a$ is the s--wave scattering length. 
This scattering length is supposed to be positive for 
$^{87}Rb$ and $^{23}Na$ but negative for $^{7}Li$. 
This means that for $^{87}Rb$ and $^{23}Na$ the interatomic 
interaction is repulsive while for $^{7}Li$ the atom--atom 
interaction is effectively attractive$^{3),10)}$. 
\par
By applying the theory of weakly interacting bosons$^{11)}$, 
the Hamiltonian operator can be written 
\beq
{\hat H} = -{\hbar^2\over 2m} \nabla^2 
+ V_{ext}({\bf r}) + V_m({\bf r})  \; ,
\eeq
where $V_{ext}({\bf r})$ is the external potential of the trap and 
$V_{m}$ is the mean--field self--consistent potential, given by 
\beq
V_m({\bf r}) = \int d{\bf r}' \; 
|\psi ({\bf r}')|^2 U({\bf r}-{\bf r}') \; , 
\eeq
where $\psi ({\bf r})$ is the wave--function of the BEC. 
Although the actual experimental traps are anisotropic$^{1),2),3)}$, it is 
simplest to consider an isotropic harmonic trap (the effect of the anisotropy 
can be treated in perturbation theory). 
The bosons are alkali atoms in the trap 
\beq
V_{ext}({\bf r})={m\omega^2\over 2}r^2 \; ,
\eeq
with zero--range effective interaction, thus by using the equation (1) we get 
\beq
V_m({\bf r}) = B |\psi ({\bf r})|^2 \; .
\eeq
\par
The mean energy of the system is given by the Gross--Pitaevskii 
functional$^{12)}$
\beq
K[\psi ] = \int d{\bf r} \; \psi^*({\bf r}) {\hat H} \psi ({\bf r}) \; , 
\eeq
and we can study the ground state stability 
by imposing the minimization of the energy functional 
\beq
\delta K[\psi ] = 0 \; .
\eeq
\par
The main point of this paper is the choice of the trial wave--function 
for the energy functional. We choose a Gaussian 
wave--function with a free parameter $\sigma$, which is 
the standard deviation of the Gaussian, i.e. the mean radius of the condensate. 
In fact, for $\sigma = \sqrt{\hbar\over m\omega}$ the test function 
is the ground--state function of the non--interacting system. 
Moreover, previous numerical results show that a Gaussian is a good 
approximation of the true ground--state function of the BEC$^{4),5),6)}$. 
\par
Let us start with the one--dimensional case. 
We choose the following trial wave--function 
\beq
\psi (x) = C \exp{({-x^2\over 2\sigma^2})} \; ,
\eeq
with the normalization condition
\beq
\int dx \; |\psi (x)|^2 = N \; ,
\eeq
where $N$ is the number of bosons, from which we obtain
\beq
C^2 = {N\over \pi^{1/2} \sigma} \; .
\eeq
By inserting this trial wave--function in the energy functional, 
after some simple calculations, we find
\beq
K={1\over 2}({\hbar^2 \over 2m})N {1\over \sigma^2} 
+{1\over 2}({m\omega^2\over 2})N \sigma^2 + 
{BN^2\over (2\pi )^{1/2}}{1\over \sigma} \; . 
\eeq
The minimum of the energy functional with respect to 
the standard deviation $\sigma$ 
is obtained by imposing the following condition 
\beq
0={dK\over d\sigma}=-({\hbar^2\over 2m})N {1\over \sigma^3} 
+({m\omega^2\over 2})N\sigma - 
{BN^2\over (2\pi )^{1/2}}{1\over \sigma^2} \; ,
\eeq
from which we get the formula 
\beq 
N={(2\pi )^{1/2}\over B}[ ({m\omega^2\over 2}) \sigma^3 - 
({\hbar^2 \over 2 m}){1\over \sigma}] \; . 
\eeq
It is easy to see that for any $B$ there is only 
one solution of the equation ${dK\over d\sigma}=0$. 
The second derivative $d^2K\over d\sigma^2$ is positive 
when ${dK\over d\sigma}=0$, namely  
the solution is stable, i.e. a minimum of the energy functional. 
In particular, for $N=0$ we have 
$\sigma = \sqrt{\hbar\over m\omega}$ and if $B>0$ 
then $\sigma \to \infty$ for $N \to \infty$, while if $B<0$ 
then $\sigma \to 0$ for $N \to \infty$. 
\par
Now we consider the three--dimensional case. 
We choose again a Gaussian trial wave--function
\beq
\psi ({\bf r}) = C \exp{({-r^2\over 2\sigma^2})} \; ,
\eeq
with the normalization condition
\beq
\int d{\bf r} \; |\psi ({\bf r})|^2 = N \; ,
\eeq
from which we find
\beq
C^2 = {N\over \pi^{3/2} \sigma^3} \; .
\eeq
The resulting energy functional is slightly different from the 
one--dimensional one
\beq
K={3\over 2}({\hbar^2 \over 2m})N {1\over \sigma^2} 
+{3\over 2}({m\omega^2\over 2})N \sigma^2 + 
{BN^2\over (2\pi )^{3/2}}{1\over \sigma^3} \; .
\eeq
We find the minimum of the energy functional with respect to 
the mean radius $\sigma$ 
by imposing the following condition
\beq
0={dK\over d\sigma}=-3({\hbar^2\over 2m})N {1\over \sigma^3} 
+3({m\omega^2\over 2})N\sigma -3 
{BN^2\over (2\pi )^{1/2}}{1\over \sigma^4} \; ,
\eeq
from which we obtain the formula
\beq
N= {(2\pi )^{3/2}\over B}[ ({m\omega^2\over 2}) \sigma^5 - 
({\hbar^2 \over 2 m})\sigma] \; . 
\eeq
By studying the function ${dK\over d\sigma}$, 
we observe that for $B>0$ there is only one solution 
(intersection with the $\sigma$--axis), which is 
a minimum of the energy functional (stable solution), instead for $B<0$ there 
are two solutions: one is a maximum (unstable solution) 
and the other a minimum (stable solution) of the energy functional. 
\par
It follows that for $B>0$, when $N=0$ we have $\sigma = 
\sqrt{\hbar\over m\omega}$ and $\sigma \to \infty$ for $N \to \infty$. 
For $B<0$ the mean radius $\sigma$ decreases by increasing the number of bosons 
$N$ to a minimum radius $\sigma^{min}$, 
with a maximum number $N^{max}$ of bosons. For greater values of $N$ 
the condensate becomes unstable and there is the collapse 
of the wave--function. It is not difficult 
to obtain the critical number of bosons. We put ${\tilde B}=-B$ and get 
\beq
0={dN\over d\sigma}={(2\pi )^{3/2}\over {\tilde B}}
[({\hbar^2 \over 2 m}) - 5({m\omega^2\over 2}) \sigma^4] \; ,
\eeq
from which we have the minimum radius
\beq
\sigma^{min}={1\over 5^{1/4}} \sqrt{\hbar \over m\omega} \; , 
\eeq
and the maximum number of bosons
\beq
N^{max} = {4 \over 5^{5/4}} {(2\pi )^{3/2}\over {\tilde B}} 
{\hbar^2\over 2m} \sqrt{\hbar\over m\omega} \; .
\eeq 
Thus we obtain an analytical formula of the maximum number 
$N^{max}$ of bosons 
for which the condensate, with attractive interaction, is stable. 
We have seen that in the one--dimensional case $N^{max}=\infty$ 
and $\sigma^{min}=0$. 
\par
For $^{7}Li$ the scattering length is $a=-14.5$ \AA , 
the axial frequency of the trap is $\omega_a/(2\pi )= 117$ Hz 
and the transverse frequency is $\omega_t/(2\pi )\simeq 163$ Hz 
(see Ref. 3 for further details). We estimate the critical 
number of particle by using $\omega /(2\pi )\simeq 120$ Hz for 
the frequency of our isotropic trap: we find $N^{max} \simeq 1400$. 
This result is in good agreement with the numerical calculations 
of Dalfovo and Stringari$^{6)}$. It is important to observe the 
the number of particles in the BEC reported in the experimental 
work of Ref. 3 is an order 
of magnitude higher than our critical value. As suggested by 
Dalfovo and Stringari$^{6)}$, the discrepancy between the experimental 
finding of Ref. 3 and the predictions of the Gross--Pitaevskii theory 
could be significantly reduced if one assumes the existence of 
a vortex in the atomic cloud$^{11)}$. 
\par
In conclusion, we have studied in the mean--field approximation 
the ground--state stability of a gas of weakly interacting bosons 
in a harmonic trap with a zero--range interaction. To minimize 
the energy functional, we have chosen a Gaussian trial wave--function, 
where its standard deviation represents the mean radius of the condensate. 
In the case of repulsive interaction the mean radius of the 
condensate grows by increasing the number of bosons. 
In the case of attractive interaction 
the mean radius decreases by increasing the number of bosons but in 
different ways depending on the dimensionality. Our results suggest 
that one must be very careful when tries to simplify a many--body problem 
by reducing its dimension. 
\par
Our technique can be applied also to bosons with non--local interaction 
and to systems with different shapes of the trapping potential$^{13)}$. 
In the future will be interesting to investigate analytically 
the collective excitation of the condensate, at least 
in the semiclassical approximation$^{14),15)}$. 

\section*{Acknowledgments}
\par
The author is grateful to Prof. A. Parola and Prof. L. Reatto 
for stimulating discussions. 

\newpage

\section*{References}

\begin{description}

\item{\ 1.} M.H. Anderson, J.R. Ensher, M.R. Matthews, C.E. Wieman, 
and E.A. Cornell, Science {\bf 269}, 189 (1995). 

\item{\ 2.} K.B. Davis, M.O. Mewes, M.R. Andrews, N.J. van Druten, 
D.S. Drufee, D.M. Kurn, and W. Ketterle, Phys. Rev. Lett. {\bf 75}, 
3969 (1995).

\item{\ 3.} C.C. Bradley, C.A. Sackett, J.J. Tollet, and R.G. Hulet, 
Phys. Rev. Lett. {\bf 75}, 1687 (1995). 

\item{\ 4.} M. Edwards and K. Burnett, Phys. Rev. A {\bf 51}, 1382 (1995).

\item{\ 5.} M. Lewenstein and L. You, Phys. Rev. A {\bf 53}, 909 (1996).

\item{\ 6.} F. Dalfovo and S. Stringari, Phys. Rev. A {\bf 53}, 2477 (1996).

\item{\ 7.} A.L. Fetter, Phys. Rev. A {\bf 53}, 4245 (1996).

\item{\ 8.} S. Stringari, Phys. Rev. Lett. {\bf 77}, 2360 (1996).

\item{\ 9.} A. Smerzi and S. Fantoni, Phys. Rev. Lett. {\bf 78}, 3589 
(1997). 

\item{\ 10.} C.C. Bradley, C.A. Sackett, and R.G. Hulet, 
Phys. Rev. Lett. {\bf 78}, 985 (1997). 

\item{\ 11.} A.L. Fetter and J.D. Walecka, {\it Quantum Theory 
of Many--Particle Systems} (McGraw--Hill, New York, 1971).

\item{\ 12.} E.P. Gross, Nuovo Cimento {\bf 20}, 454 (1961); 
L.P. Pitaevskii, Sov. Phys. JETP {\bf 13}, 451 (1961).

\item{\ 13.} A. Parola, L. Salasnich and L. Reatto, 
"Tuning the interaction strangth in bosonic clouds: How to 
use Alkali atoms with negative scattering length", 
submitted to Phys. Rev. A. 

\item{\ 14.} M. Robnik and L. Salasnich, J. Phys. A {\bf 30}, 
1711 (1997); M. Robnik and L. Salasnich, J. Phys. A {\bf 30}, 1719 (1997). 

\item{\ 15.} L. Salasnich, Mod. Phys. Lett. B {\bf 11}, 269 (1997). 

\end{description}

\end{document}